\documentclass[twocolumn,epsf,showpacs,prd]{revtex4}
\usepackage{graphicx}% Include figure files
\usepackage{dcolumn}% Align table columns on decimal point
\usepackage{bm}% bold math
\usepackage{amsmath}
\begin{document}

\newcommand{\be}{\begin{equation}}
\newcommand{\ee}{\end{equation}}
\newcommand{\ba}{\begin{eqnarray}}
\newcommand{\ea}{\end{eqnarray}}
\def\bone{$B^{(1)}$}
\def\bone{B^{(1)}}
\def\etal{{\it et al.~}}
\def\eg{{\it e.g.~}}
\def\ie{{\it i.e.~}}
\def\DM{dark matter~}
\def\DE{dark energy~} 
\def\GC{Galactic centre~} 
\def\susy{SUSY~}

\title{Neutrinos from Dark Matter annihilations at the 
Galactic Centre}

\author{Gianfranco Bertone$^a$, Emmanuel Nezri$^b$, Jean Orloff$^c$, Joseph Silk$^{d}$}
\affiliation{$^a$ NASA/Fermilab Theoretical Astrophysics Group, 60510 Batavia IL\\
$^b$ Laboratoire de Physique Th\'{e}orique des Hautes Energies Universit\'{e}
Paris-Sud, F-91405 Orsay\\ 
$^c$ Laboratoire de Physique Corpusculaire, Universit\'{e} Blaise Pascal, F-63177 Aubi\'{e}re\\
$^d$ Astrophysics, Denys Wilkinson Building, Keble Road, Oxford OX1 3RH, UK}

\vspace{0.5truecm}
\begin{abstract}
We discuss the prospects for  detection of high energy neutrinos 
from dark matter annihilation at the Galactic centre. 
Despite the large uncertainties associated with our  poor 
knowledge of the distribution of dark matter in the innermost 
regions of the Galaxy, we determine an upper limit on the neutrino flux 
by requiring that the associated gamma-ray emission does not exceed the 
observed flux. We conclude that if dark matter is made of neutralinos, 
a  neutrino flux from dark matter annihilations at the GC will not be
observable 
by Antares. Conversely, the positive detection of such a flux would either 
require an alternative explanation, in terms of astrophysical processes, 
or the adoption of other \DM candidates, disfavouring the case for
neutralinos.
\end{abstract}

\pacs{... \hspace{5cm} FERMILAB-Pub-04/032-A, LPT Orsay-04/22}
\maketitle

\vspace{1truecm}

\section{Introduction}

There is robust observational evidence for the dominance of
non-baryonic
dark matter 
over baryonic matter in the universe. Such evidence comes from many independent
observations over different length scales. The most stringent 
constraint on the abundance of dark matter comes from the 
analysis of CMB anisotropies. In particular, the WMAP experiment 
restricts  the abundance of matter to lie in the range 
$\Omega_M h^2=0.135^{+0.008}_{-0.009}$~\cite{Spergel:2003cb}.
The same type of analysis constrains the amount of baryonic 
matter to be in the range $ \Omega_b h^2= 0.0224\pm 0.0009$,
in good agreement with predictions from Big Bang nucleosynthesis 
$0.018 < \Omega_b h^2 < 0.023$ (e.g. Ref.~\cite{Olive:2003iq}).

It is commonly believed that such a non-baryonic component 
could consist  of new, as yet undiscovered, particles, usually  referred to
as WIMPs (Weakly Interacting Massive Particles). 
It is intriguing that some extensions of the standard 
model of particle physics predict the existence of
particles that would be excellent DM candidates.
In particular great attention has been recently devoted to 
candidates arising in supersymmetric theories. The lightest 
supersymmetric particle (LSP), which in most supersymmetric
scenarios is the so--called {\it neutralino}, is stable in 
theories with conservation of R--parity, and can have masses 
and cross sections of typical WIMPs.

One possible way of probing the nature of dark matter particles 
is to look for their annihilation signal \cite{ss:1984}. For this purpose, 
the best regions to examine  are those where the dark matter
accumulates, the annihilation rate being proportional to
the square of the particle number density. A wide literature
exists discussing the prospects of observing annihilation 
radiation from the Galactic centre (e.g. Refs.~\cite{Berezinsky:1994wv,
Bergstrom:1997fj, Bertone:2002je}), high energy 
neutrinos from the Sun (e.g. Refs.~\cite{sos:1985, Bertin:2002ky, Bergstrom:1998xh}), 
gamma-rays and synchrotron from dark matter clumps in the galactic halo 
(e.g. Refs.~\cite{Bergstrom:1998zs, Blasi:2002ct, Tasitsiomi:2002vh, Stoehr:2003hf}), 
gamma-rays from external galaxies (e.g. Refs.~\cite{Baltz:1999ra, Falvard:2002ny, 
Tasitsiomi:2003vw, Pieri:2003cq}), positrons and 
antiproton (e.g. Refs.~\cite{Chardonnet:1996ca, Bergstrom:1999jc,
Baltz:2001ir, Donato:2003xg}) and more.

Large uncertainties are associated with predictions of  
annihilation fluxes, due to our  poor knowledge of the distribution
of dark matter, especially in the innermost regions of the 
Galaxy. Numerical simulations suggest that the dark matter
density is well approximated by ``cuspy'' profiles, with 
a power-law behaviour $\rho \propto r^{-\gamma}$.
Estimates of $\gamma$ vary between having no cusp, $\gamma \sim 0,$
\cite{be:2001}, to a cusp $\gamma =1$ that is further steepened by
adiabatic compression of the baryons \cite{prada:2004}.
One can trace these differences in large part to uncertainties in the
stellar mass in the inner galaxy as inferred from microlensing experiments.
The poor knowledge of $\gamma$ implies uncertainties of
several  orders of magnitude in the annihilation flux. The
situation is made even worse by the possible influence on
the dark matter profile of the  probable adiabatic formation of
the supermassive black hole 
lying at the Galactic centre.
Such uncertainties make indirect searches less effective 
for constraining  the physical parameters (such as mass and cross 
sections) of dark matter particles.  
 
We suggest here a method for  evading the  astrophysical 
uncertainties in the neutrino flux, by requiring that 
the associated gamma-ray emission does not exceed the 
flux observed by the EGRET experiment in the direction 
of the Galactic centre. In fact, if we normalize the
gamma-ray flux to the EGRET data, the corresponding 
neutrino flux will be an {\it upper limit} on the 
actual neutrino flux measurable on Earth. Choosing 
the EGRET normalization corresponds to fixing the product 
$J\sigma v N_{\gamma}$, where the quantity J, defined below, 
includes all of the astrophysical information,  
$\sigma v$ is the total annihilation cross section 
and $N_{\gamma}$ is the number of photons produced per annihilation.

This  paper is organised as follows: we first discuss the
gamma--ray source observed by the EGRET satellite in the
direction of the \GC; in Sec. III we briefly review the results
on the distribution of \DM from observations and N-body simulations.
In Sec. IV we present the particle physics details of our candidate,
the neutralino, arising in supersymmetric theories, in Sec. V 
we review the prospects of indirect detection of such candidates 
through gamma--ray and neutrino emission, for a typical \DM profile,
and in Sec. VI we compare the prospects of indirect detection through
annihilation radiation from the GC with other searches. 
We present in Sec. VII the upper limit on the neutrino flux,
obtained by normalizing the annihilation flux to the EGRET
data, and we finally give our conclusions in Sec. VIII.

\section{The EGRET source at the Galactic centre}

The Galactic centre region has been observed by 
EGRET, the Energetic Gamma Ray Experiment Telescope, 
launched on the Compton Gamma Ray Observatory in 1991,
and sensitive to an energy range 30MeV--30GeV. A strong
excess of emission was observed in an error circle of 
0.2 degree radius including the position $l=0^\circ$, 
$b=0^\circ$, the strongest emission maximum lying within 15 
degrees from the GC~\cite{mayer}.

The radiation exceeds, and also is harder than, the expected gamma ray emission 
due to the interaction of primary cosmic rays with the interstellar
medium (see \eg Strong \etal 1998~\cite{Strong:1998fr}).
At the energies we are interested in, $E \gtrsim 1$~GeV,
the main source of photons is the decay of $\pi^0$ mesons
originating from processes such as 
\be
p \;\; + \;\; X \;\; \rightarrow \;\; \pi^0
\nonumber
\ee
\be
He \;\; + \;\; X \;\; \rightarrow \;\; \pi^0
\nonumber
\ee
where X is an interstellar atom. The interested reader will
find a detailed estimate of the background radiation in 
Cesarini \etal 2003~\cite{Cesarini:2003nr}.

It is intriguing to conjecture that such excess emission 
could originate from \DM annihilation at the \GC. However, 
such an interpretation is problematic. In fact, as noticed 
by Hooper and Dingus ~\cite{Hooper:2002ru}, the EGRET source
 is not exactly coincident with the \GC, which
would make the interpretation of the signal as due 
to the annihilation in a spike around the Galactic centre 
at least problematic.

Furthermore there is some evidence, although weak, that
the source could be variable. Such a result could rule out
completely the interpretation of the excess emission as
due to annihilation radiation from the \GC. The 
variability of 3EG J1746-2851 has been recently 
discussed in Nolan \etal 2003~\cite{nolan}.
An additional flaw has been pointed out by P.Salati
\footnote{P.Salati, 2003, private communication}, 
namely the fact that the HI 
column density was merely interpolated in the region 
of interest, where it was thought to be unreliable due
to strong self-absorption and high optical thickness. 
It is an open question how the conclusions would 
change if different assumptions are made about  the HI 
column density.

Here, we will regard the EGRET observation as an 
{\it upper limit} on the annihilation gamma-ray flux from the Galactic 
centre.

\section{Dark matter distribution}

The usual parametrization for \DM density profiles is
\begin{equation}
  \rho(r)= \frac{\rho_0}{(r/R)^{\gamma}
  [1+(r/R)^{\alpha}]^{(\beta-\gamma)/\alpha}} \;\;.
\label{profile} 
\end{equation}
where $r$ is the galacto-centric coordinate, $R$ is a 
characteristic length and $\alpha, \beta$ and $\gamma$ are 
free parameters. 

There is consensus, at present, about the shape
of the profile in the outer parts of  halos, but not in the
innermost regions, due to loss of numerical resolution in
N-body simulations and to the poor resolution in 
observation of rotation curves of outer galaxes.
 Navarro, Frenk \& White \cite{Navarro:1995iw}, found with
N-body simulations that the profile could be well approximated 
at small radii with a power-law $\rho(r)=r^\gamma$ with 
$\gamma \sim 1$. Other groups reached different 
conclusions (see e.g. Refs.~\cite{Moore:1999gc, kra}).

The most recent N-body simulations \cite{Hayashi:2003sj, fuk:2003, stoehr:2004} suggest 
that profiles do not approach power laws with a well-defined 
index at very small radii. Profiles continue to become shallower,
i.e. the (negative) logarithmic slope becomes higher, when moving 
towards the centre. Some  authors however contend that convergence is
reached with $\gamma \approx 0.2-0.3$ at $0.3\%$ of the virial radius \cite{diemand:2004}.

An additional complication of the \DM profile at the center
of our Galaxy, is the well-established presence of a a $3.6\times 10^6$ 
solar mass black hole (see e.g.Ref.~\cite{schodel}), that would accrete 
dark matter, producing a so-called 'spike'~\cite{Gondolo:1999ef}, and leading to
an enhancement of the annihilation flux by several orders of 
magnitude (see  Ref.~\cite{Bertone:2002je}, and references therein, 
for a discussion of indirect detection of \DM in presence of spikes, 
and of dynamical effects that could potentially destroy them).

The observational situation is even less clear. The analysis of
rotation curves of galaxies has led some authors to claim 
inconsistency
of the observed 'flat' profiles with the cuspy profiles 
predicted by N-body simulations. Other groups \cite{Swaters:2002rx,vandenBosch:1999ka} 
claim instead that cuspy profiles are  
compatible with observations. 
Hayashi et al \cite{Hayashi:2003sj} compared the observational data
directly with 
their numerical simulations (rather than fits of 
their simulations) and found no significant discrepancy
in most cases. They attributed the remaining discrepancies
to the difference between circular velocities and gas rotation
speed in realistic triaxial halos.

It is clear that the predictions of annihilation fluxes are
strongly affected by the uncertainties in \DM distribution. 
In particular, since the annihilation rate is proportional 
to the square of the particle density, different profiles can
lead to uncertainties of many orders of magnitude.
To get around these, we will use the gamma--ray flux observed by EGRET in the
direction of the \GC, to get rid of astrophysical uncertainties and 
produce a robust upper limit on the neutrino flux from 
\DM annihilation at the \GC.

\section{Supersymmetric DM: neutralinos}

Neutralinos are by far the best studied \DM candidates. 
They arise in supersymmetric theories with conservation
of R-parity, in which the lightest supersymmetric particle 
(LSP) cannot decay in standard model particles, and is thus stable. 
In most cases the LSP is the neutralino, i.e. a linear 
combination of the supersymmetric partners of the gauge and higgs bosons
\be
\chi (\equiv \tilde{\chi^0_1})= z_{11} \tilde B+  z_{12}\tilde W_3 + z_{13} \tilde H_1^0 + z_{14} \tilde H_2^0.
\ee
The matrix $z$ diagonalizes the neutralino mass matrix, which is expressed as
\begin{equation} \label{eq:neumass}
%  {\cal M}_{\tilde \chi^0} = 
%  {\cal M} = 
  \left( \begin{array}{cccc}
M_1 & 0 & -m_Z c_{\beta} s_{W} & m_Z s_{\beta} s_{W}\\
0 & M_2 & m_Z c_{\beta} c_{W} & -m_Z s_{\beta} c_{W}\\
-m_Z c_{\beta} s_{W} & m_Z c_{\beta} c_{W} & 0 & -\mu\\
m_Z s_{\beta} s_{W} & -m_Z s_{\beta} c_{W} & -\mu & 0
  \end{array} \right)
\end{equation}
in the basis $(\tilde{B},\tilde{W}_{3},\tilde{H}_{1}^0,\tilde{H}_{2}^0)$.

Similarly defining $V_{11(2)}$ as the wino (higgsino) fraction of the lightest
chargino, the neutralino annihilation channels and cross-sections most relevant
for indirect detection are
\begin{equation}
\begin{array}{ll}
  \chi \chi \xrightarrow{A}b \bar{b}&:\ 
  \sigma \propto [z_{11(2)}z_{13(4)}]^2\\
  \chi \chi \xrightarrow{Z}Z h &:\ 
  \sigma \propto [z^2_{13(4)}]^2\\
  \chi \chi \xrightarrow{\chi^+}W^+W^-  &:\ 
  \sigma \propto [z_{13(4)}V_{12}]^2  \textrm{ and/or }  [z_{12}V_{11}]^2
\end{array}
\end{equation}
Annihilation in these channels thus increases with the wino or higgsino fraction
of the neutralino. The spectra of the indirect detection signals studied
here keep an imprint of the dominant channel.

\paragraph*{For muon via neutrino production:}
the $W^+W^-$ and $Z h$ channels produce more energetic neutrinos, {\it i.e} a
harder neutrino spectrum than $b \bar{b}$. Both the neutrino-nucleon cross
section ($\sigma_{\nu-N}$) and the muon range ($R_{\mu}$) being proportional
to neutrino energy, harder spectra give higher muon detection rates for the
threshold considered here (5 GeV): $\phi_{\mu} \propto \phi_{\nu}
\sigma_{\nu-N}(E_{\nu}) R_{\mu}(E_{\nu})$.

\paragraph*{For gamma production:}
the $b\bar{b}$ and also $t\bar{t}$ channels dominate the $\gamma$
spectra around 2 GeV but at higher energies, the harder $WW$ and $Zh$
channels come in. Experiments with different thresholds can thus see
different processes.

The influence of the dominant annihilation channel is displayed on
figure \ref{NeutrinosEGRET} below.

We have performed a scan of SUSY models at the GUT scale, computing
renormalisation group equations and radiative electroweak symmetry breaking
with {\tt Suspect} \cite{Suspect}, the neutralino relic density with {\tt
  Micromegas} \cite{Micromegas} and detection rates with {\tt Darksusy}
\cite{Darksusy} \footnote{The correction in the latest version Darksusy 4
  would reduce all absolute fluxes reported here by a factor of 2, but our
  main results on ratios between different fluxes or fluxes normalized to
  EGRET are insensitive to this correction.}. The SUSY models explored fall
in 3 classes (see \cite{Bertin:2002ky} for definitions)

\paragraph*{CMSSM:} with universal scalar $m_0$ and  
gaugino $m_{1/2}$ mass parameters in the ranges:\\
$50\textrm{GeV}<m_0<4000\textrm{GeV}$,
$50\textrm{GeV}<m_{1/2}<2000\textrm{GeV}$, $A_0=0$, $\tan{\beta}=5,20,35$

\paragraph*{Non universal gaugino mass $M_2|_{GUT}$:} same values as above,
except for $M_2|_{GUT}=0.6 m_{1/2}$ (instead of $1m_{1/2}$), leading to
$M_2 \sim M_1$ in the neutralino mass matrix (eq. \ref{eq:neumass}); the resulting
non-zero wino contents ($z_{12}$) allows for non-negligible relic
densities.

\paragraph*{Non universal gaugino mass $M_3|_{GUT}$:} same values as in
the universal case ($\tan{\beta}=20,35$ only) with $M_3|_{GUT}=0.6 m_{1/2}$
(instead of $1m_{1/2}$), to decrease the $\mu$ parameter in the neutralino
mass matrix (eq. \ref{eq:neumass}) to favour the higgsino fraction
($z_{13(4)}$) and decrease scalar masses, in particular the pseudo scalar
$A$ mass. We do not relax Higgs sector universality, whose interesting effects on 
dark matter are similar to those of a lower $M_3|_{GUT}$.

Finally, we apply the following conservative cuts on our models:

\begin{itemize}
\item Higgs mass: $ m_h > 113.5$ GeV \cite{Higgslimit},

\item Chargino mass: $m_{\chi^+} > 103.5$  GeV \cite{charginolimit},

\item Relic density: $0.03<\Omega_{\chi}{\rm h}^2<0.3$, 
but we also show the WMAP~\cite{Spergel:2003cb} range $
\Omega^{WMAP}_{\rm CDM} {\rm h}^2 = 0.1126^{+0.0161}_{-0.0181}$,

\item $ b \rightarrow s \gamma$ Constraint  \cite{bench}:\\ 
$2.33 \times 10^{-4} < \mathrm{BR} (b \rightarrow s \gamma) < 4.15 
\times 10^{-4}$,

\item The muon anomalous magnetic moment \cite{Hagiwara:2003da}: $
  8.1~10^{-10}<\delta_{\mu}^{\mathrm{susy}} =
  \delta_{\mu}^{\mathrm{exp}}-\delta_{\mu}^{\mathrm{SM}}<44.1~10^{-10} \ 
  [2\sigma] $. 
\end{itemize}
Given the recent evolution of this last range, the on-going debate about
the use $\tau$-decay data \cite{Davier:2003pw} and the drastic effect of
this 2$\sigma$ cut, which both excludes the SM and many interesting dark
matter models, the range $0<\delta_{\mu}^{\mathrm{susy}}<8.1~10^{-10}$ 
will not be discarded, but
displayed in pale on all plots.

%\subsection{Kaluza--Klein particles}

%The study of LKP as a viable \DM candidate dates
%back to the work of Kolb \& Slansky 1984~\cite{Kolb:fm}.
%The LKP has been then reconsidered in the framework of UED,
%in which it is likely to be associated with the first KK 
%excitation of the photon, more precisely the first KK excitation 
%of the Hypercharge gauge boson \cite{Cheng:2002iz}, and we will 
%refer to it as $\bone$. 
%
%An accurate calculation of the $\bone$ relic density was
%performed by Servant \& Tait 2002~Ref.~\cite{Servant:2002aq}, 
%who found that if the LKP is to account for DM then its mass (which is 
%inversely proportional to the compactification radius $R$) should lie in 
%the range 400--1200 GeV, \ie above any current experimental 
%constraint.%
%
%The annihilation cross section has been studied in 
%Ref.~\cite{Servant:2002aq}, and can be numerically approximated
%to
%\be
%\sigma v = \frac{0.6 \mbox{pb}}{m_{\bone}^2 (\mbox{TeV})} \;\; .
%\ee
%The branching ratios for the annihilation process
%are almost independent on the particle mass. In particular we note
%that the branching raton into charged lepton pairs is $\sim 60\%$ and
%the branching ratio into neutral  lepton pairs is $\sim 4\%$.
%
%As opposite to the case of neutralinos, Kaluza--Klein particles can
%directly annihilate into neutrinos with a quite large cross section.
%The spectrum of these neutrinos would be a line at $E=m_{\bone}$,
%\ie at the maximum available energy, which makes detection in 
%neutrino telescopes easier ~\cite{bertone}.

\section{Gamma--ray and neutrino flux from the GC} 

Indirect detection of Dark Matter is based on observation of annihilation
products like gamma-rays, neutrinos or synchrotron emission of secondary
electron--positron pairs.  The spectrum of secondary particles of species
$i$ from annihilation of DM particles whose distribution follows a profile
$\rho(r)$ where $r$ is the Galacto--centric coordinate, is given by
\begin{equation}
\Phi_i(\psi,E)=\sigma v \frac{dN_i}{dE} \frac{1}{4 \pi M^2}
\int_{\mbox{line of sight}}d\,s
\rho^2\left(r(s,\psi)\right)\label{flux}
\end{equation}
where the coordinate $s$ runs along the line of sight, in a
direction making an angle $\psi$ respect to the direction
of the GC. $\sigma v$ and $dN_i/dE$ are respectively the 
annihilation cross section and the spectrum of secondary particles
per annihilation, while $M$ is the mass of the annihilating
DM particle.

To isolate the factor depending on astrophysics, i.e. 
the integral of $\rho^2$ along the line of sight, we introduce,
following \cite{Bergstrom:1997fj}, the quantity $J(\psi)$
\begin{equation}
J\left(\psi\right) = \frac{1} {8.5\, \rm{kpc}} 
\left(\frac{1}{0.3\, \mbox{\small{GeV/cm}}^3}\right)^2
\int_{\mbox{\small{line of sight}}}d\,s\rho^2\left(r(s,\psi)\right)\,.
\end{equation}
and its average over a spherical region of solid angle 
$\Delta\Omega$, centered on $\psi=0$, $\overline{J}(\Delta\Omega)$.

With these definitions the flux from a solid angle $\Delta\Omega$ is
\ba
\nonumber
\Phi_{i}(\Delta\Omega, E)\simeq5.6\times10^{-12}\frac{dN_i}{dE} 
&\left( \frac{\sigma v}
{\rm{pb}}\right)\left( \frac{1\rm{TeV}} 
{\rm{M}}\right)^2 \overline{J}\left(\Delta\Omega\right) \\
& \times \; \Delta\Omega\,\rm{cm}^{-2} \rm{s}^{-1}\,.
\label{final}
\ea

Apart from astrophysics, large uncertainties on the
quantities in eq.~\ref{final} are associated with
the details of particle physics. The dependence of
the annihilation cross section on the mass $M_\chi$
is different for each DM candidate, and even in the
framework of a specific supersymmetric scenario,
cross sections for a given mass could span over
several orders of magnitude.

%\section{Gamma--ray and neutrino flux}
\begin{figure}[h]
\includegraphics[width=0.48\textwidth,clip=true]{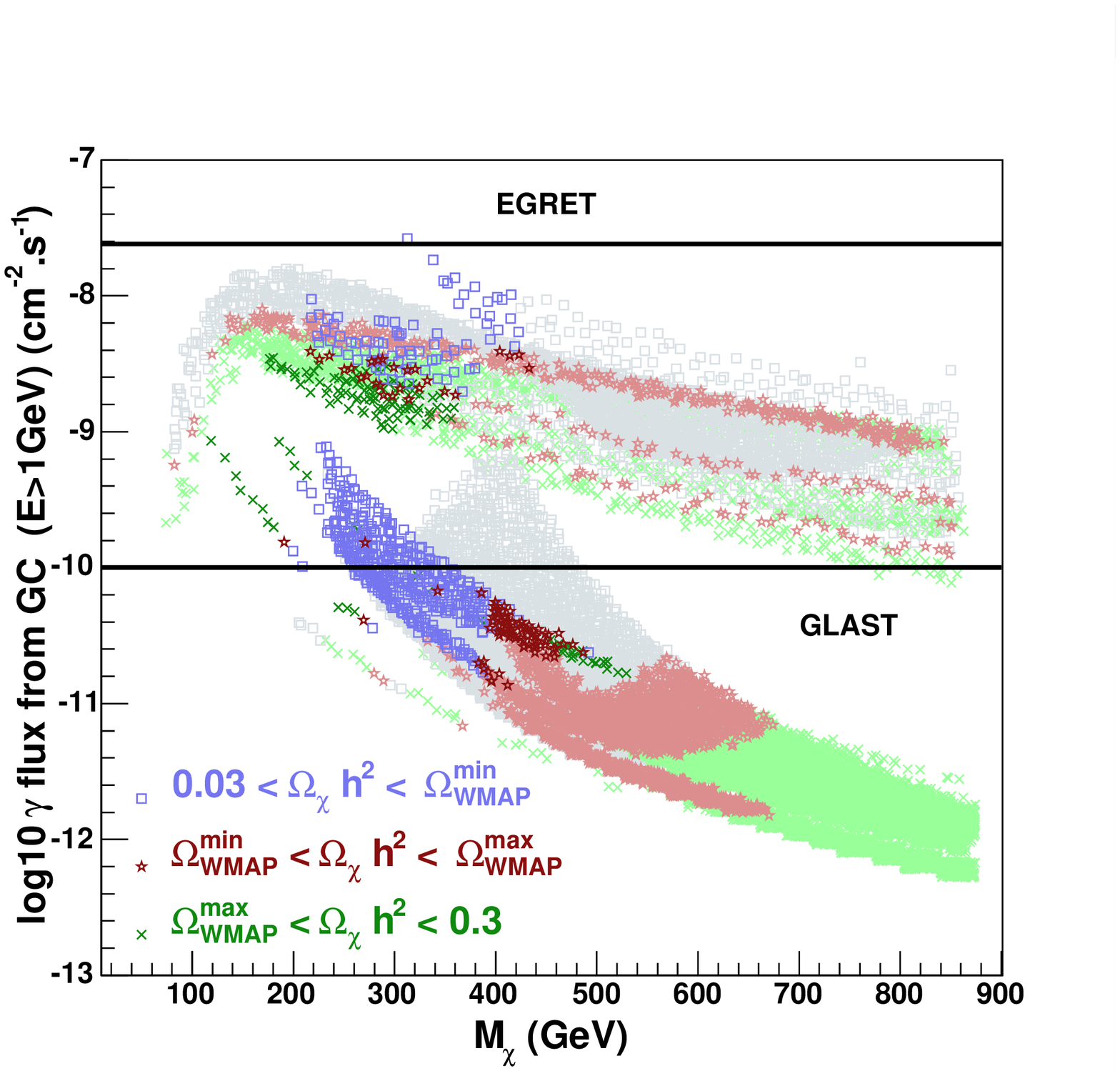}
\caption{Gamma--ray flux from neutralino annihilation at the GC, 
assuming a NFW profile. For comparison we show the EGRET and GLAST 
sensitivities.
Shades paler than in the legend denote a low $\delta_\mu^{\mathrm{susy}}$ value.
}
\label{gammas}
\end{figure}

%The annihilation cross sections of DM particles and 
%the branching ratios for different channels depend
%on the specific candidate chosen.

We show in Fig.~\ref{gammas} the gamma--ray flux from 
neutralino annihilation at the \GC assuming a NFW profile,
along with EGRET and GLAST sensitivities. We see that all
the supersymmetric models predict fluxes below the EGRET 
sensitivity in this case, but many of them could produce
fluxes observable by GLAST.

Nevertheless, as already mentioned, EGRET {\it did}
observe a source at the \GC, although it is unclear 
whether this emission is actually to be attributed to 
WIMP annihilations. We adopt a conservative approach
and consider the EGRET source as an upper limit on  the
WIMP annihilation flux. 
In this sense, we see from Fig.~\ref{gammas} that
if neutralinos are the dark matter particle, then 
there is room for profiles even more ``cuspy'' than
NFW.

%\subsection{Neutrino flux}

\begin{figure}[h]
\includegraphics[width=0.48\textwidth,clip=true]{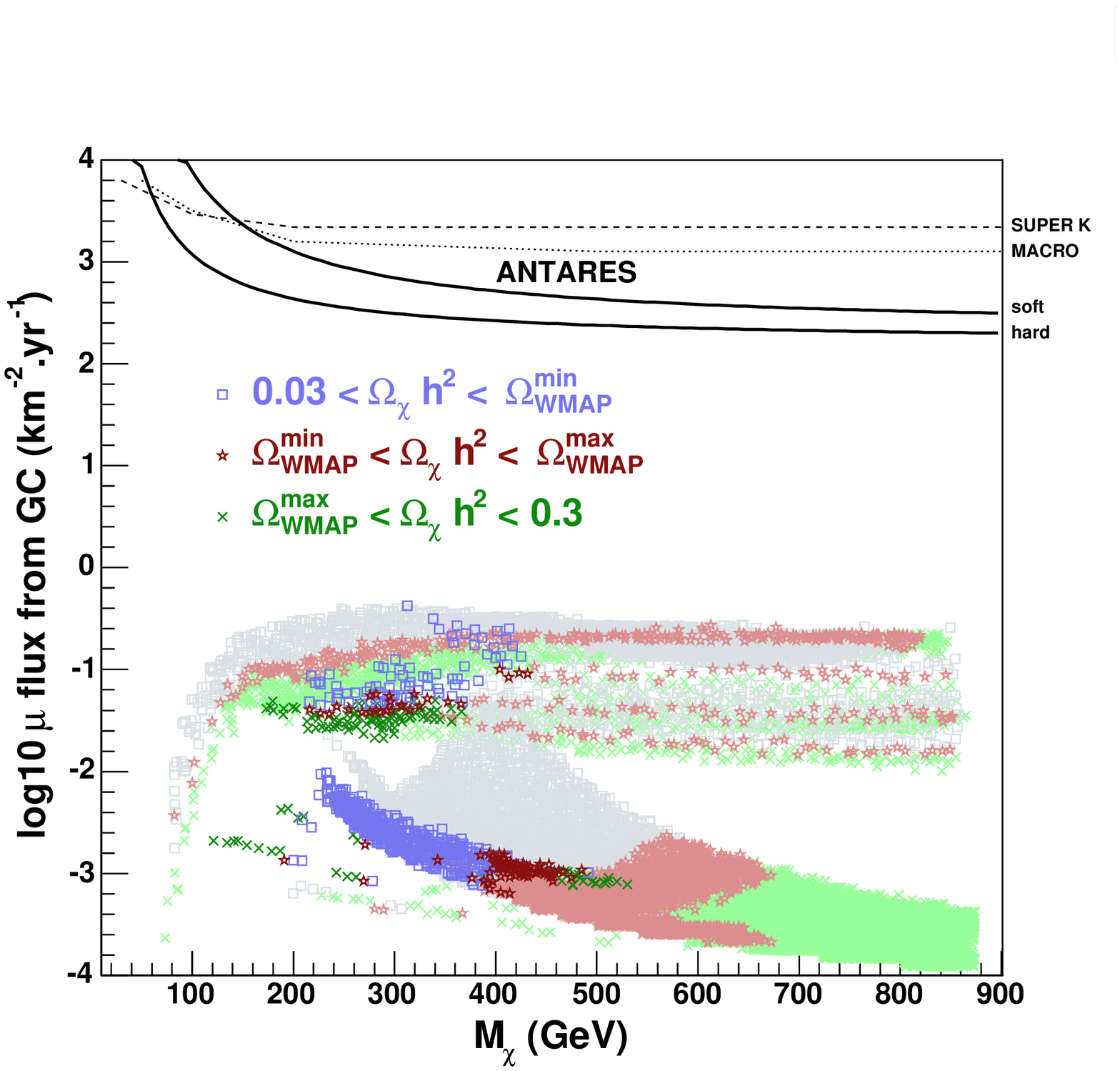}
\caption{Neutrino-induced muon flux from neutralino annihilation at the GC, 
assuming a NFW profile. For comparison we show the expected Antares 
sensitivity.
Shades paler than in the legend denote a low $\delta_\mu^{\mathrm{susy}}$ value.
}
\label{neutrinos}
\end{figure}

Always assuming a NFW profile, we show in  Fig.~\ref{neutrinos} the 
neutrino-induced muon flux from dark matter annihilation at the GC. We show for 
comparison the expected sensitivity of the Antares telescope (e.g. 
~\cite{Antares}, currently
under construction in the Mediterranean sea. The telescope sensitivity
depend on the incoming neutrino spectrum, we thus show two sensitivity 
curves (for a 3-years period of observation), one relative to a hard
flux (relevant for the $W^+W^-$ and $Z h$ channels),
the other relative to a soft flux (relevant for the  $b \bar{b}$ channel).As 
can be seen, the predictions fall  several orders of magnitude 
below the Antares sensitivity. 
Of course, at this stage, this 
does not  necessarily imply that Antares will not observe any
neutrinos from the \GC, as we have seen in the previous section
that it is possible that the actual \DM profile is steeper 
than NFW, adopted for Figs.~\ref{gammas} and ~\ref{neutrinos}. 

\section{Comparison with other searches}

In this section we compare, for completeness, 
the prospects of detection  of the SUSY models 
discussed above with other detection
techniques, 
%%% do not keep if Hess included here:
which are actually insensitive to the
profile of \DM in the innermost regions of the Galaxy.

\begin{figure}[h]
\includegraphics[width=0.48\textwidth,clip=true]{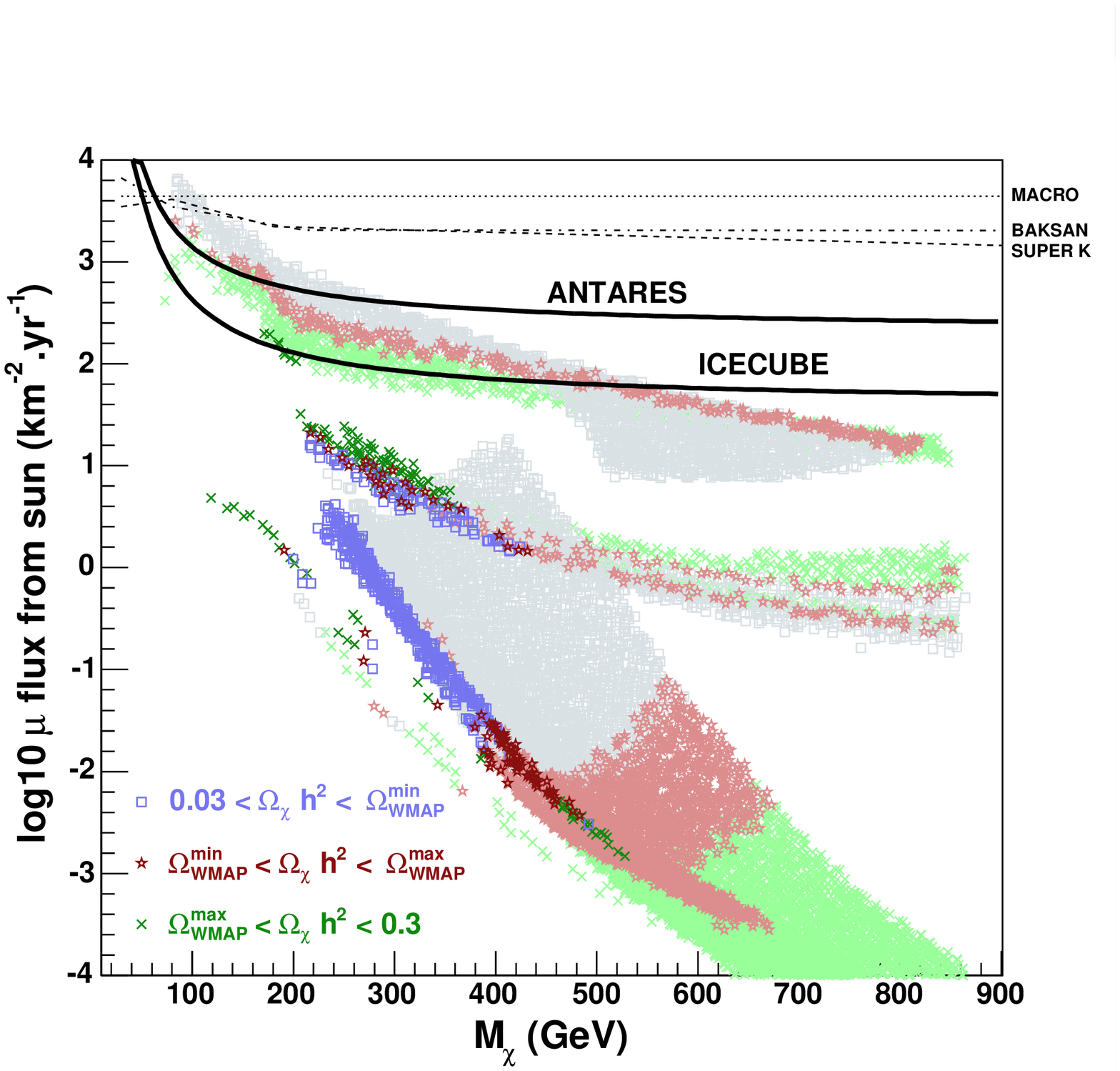}
\caption{Neutrino-induced muon flux from neutralino annihilation in the solar core.
Shades paler than in the legend denote a low $\delta_\mu^{\mathrm{susy}}$ value.
}
\label{solar}
\end{figure}

In Fig.~\ref{solar} we show the flux of neutrinos from neutralino
annihilation in the solar core. The projected sensitivities of both Antares
and IceCube ~\cite{Ice3} appear to be able to probe the 
supersymmetric models with the
non-negligible higgsino fraction necessary for an efficient
neutralino capture rate in the Sun.

\begin{figure}[h]
\includegraphics[width=0.48\textwidth,clip=true]{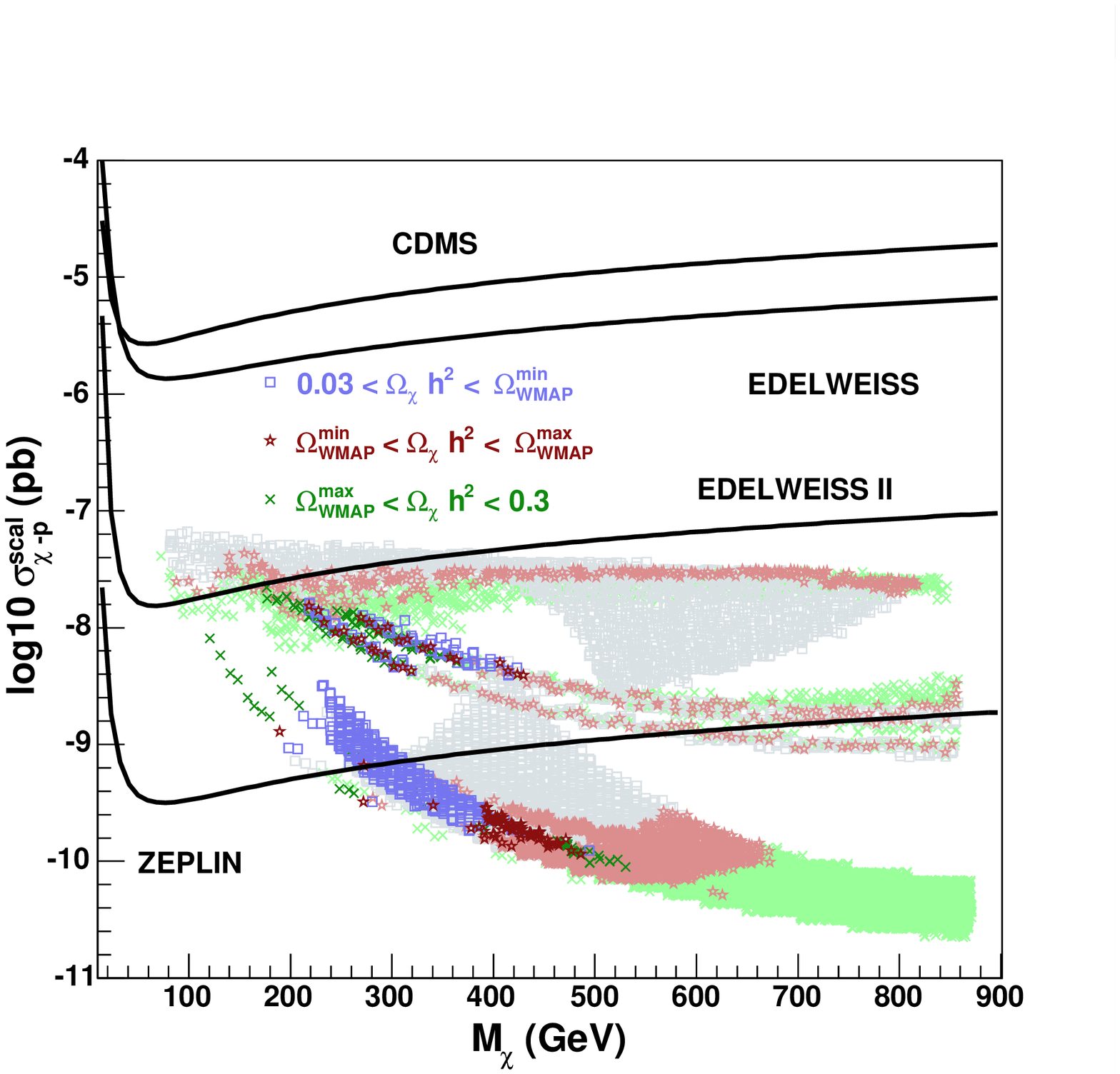}
\caption{Prospects of neutralino direct detection.
Shades paler than in the legend denote a low $\delta_\mu^{\mathrm{susy}}$ value.
}
\label{direct}
\end{figure}

We also show in Fig.~\ref{direct} the potential of direct
detection techniques to probe the neutralino nature through 
the search for neutralino-nucleon interactions in large
detectors, such as Edelweiss~\cite{Edelweiss} and CDMS
~\cite{Abrams:2002nb}. There are a couple of orders of 
magnitude between the present
-day experiment sensitivities and the most
optimistic predictions for neutralinos. But this gap could be
bridged by next-generation experiments such as  Edelweiss II 
(e.g. ~\cite{EdeII}) and Zeplin~\cite{Zeplin}.

\section{Upper limit for the Neutrino flux}

In order to maximize the neutrino flux from \DM 
annihilation at the \GC, we normalize the flux of 
gamma--rays, associated with such a  neutrino flux, to the
EGRET data. 
This corresponds to fixing, for each model, the product 
$J \, \sigma v \, N_{\gamma}$, with  
$N_{\gamma}= \sum_i N_i \, R_i$;
here  $R_i$ is the branching ratio of all the channels $i$
contributing $N_i$ gamma--rays above a given threshold energy.

Having fixed the particle physics contents of our \DM candidate,
the ratio between the number of photons and
the number of neutrinos emitted per annihilation is known.
We can thus estimate the neutrino flux from the \GC
associated with a gamma-ray emission reproducing the EGRET
data. Finally we can convert the flux of neutrinos into
a flux of muons, produced by neutrinos interactions with
the rock around detectors on Earth, in order to compare
with experimental sensitivities.

The rescaled flux of muons $\phi^{\rm{norm}}_{\mu}(>E_{th})$ 
will thus be given by
\begin{equation}
\phi^{\rm{norm}}_{\mu}(>E_{th})= \frac{\phi^{\rm{NFW}}_{\mu}(>E_{th})
\, \phi^{\rm{EGRET}}_{\gamma}(E_*)}{\phi^{\rm{NFW}}_{\gamma}(E_*)} \,
\end{equation}
where the label NFW reminds that NFW profiles have been 
used to compute  profile-independent flux ratios,
and $E_*$ is the energy at which we decide to normalize 
the flux to the gamma-ray data (in our case $E_*=2$GeV).

\begin{figure}[h]
\includegraphics[width=0.48\textwidth,clip=true]{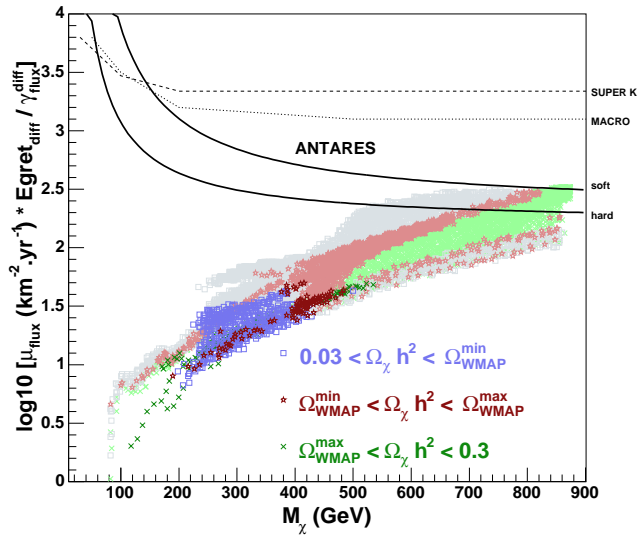}
\caption{Neutrino-induced muon flux from the \GC normalized to EGRET with relic density
  values.
Shades paler than in the legend denote a low $\delta_\mu^{\mathrm{susy}}$ value.
}
\label{NeutrinosEGRETOh2}
\end{figure}

\begin{figure}[h]
\includegraphics[width=0.48\textwidth,clip=true]{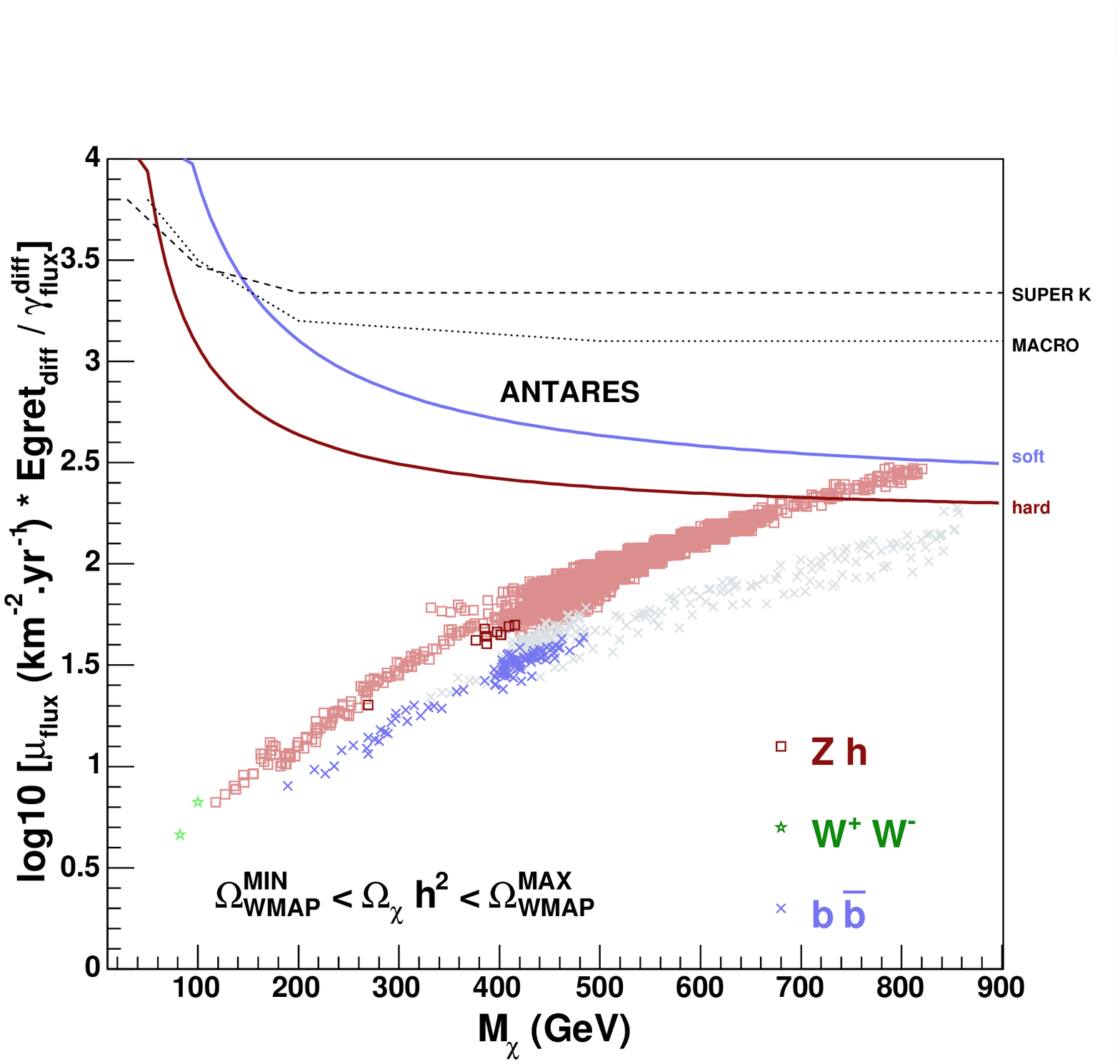}
\caption{Neutrino-induced muon flux from the \GC normalized to EGRET, for models with 
  WMAP-preferred relic density sorted by leading ($\equiv BR>0.5$)
  annihilation channel.  Shades paler than in the legend denote a low
  $\delta_\mu^{\mathrm{susy}}$ value.  }
\label{NeutrinosEGRET}
\end{figure}

The results are shown in Figs.~\ref{NeutrinosEGRETOh2} and
\ref{NeutrinosEGRET}. The
muon flux normalised to the EGRET data represent an 
upper limit, as the observed gamma--ray emission could 
be due to processes other than \DM annihilation. 
The comparison with the Antares sensitivity shows that
only the highest mass neutralinos can possibly be detected in the
Galactic centre. Insisting on the WMAP relic density in
Fig.~\ref{NeutrinosEGRET} and using the hard neutrino spectrum sensitivity
appropriate to the relevant $Zh$ channel, we need at least 700~GeV
neutralinos, whose contribution to the muon anomalous moment is similar to
the (excluded?) Standard Model.

If neutrinos are nevertheless observed above the given fluxes, then their
interpretation  as due to neutralino annihilation is problematic and
would actually require either the adoption of other \DM candidates
annihilating dominantly into neutrino pairs or a different explanation,
e.g. in terms of astrophysical sources.

Concerning other \DM candidates, a case-by-case 
analysis is needed. For Kaluza--Klein candidates (e.g. 
Ref.~\cite{Servant:2002aq} and references therein) , 
in particular, there are several channels 
contributing to the neutrino flux (see Ref.~\cite{Bertone:2002ms}).
Neutrinos coming from the decay of charged pions originating
in quark fragmentations have a relatively soft spectrum, 
and cannot be detected with Antares, even normalizing the 
gamma--ray flux to the EGRET data. A similar conclusion
applies for neutrinos from prompt semi-leptonic decay of 
secondary heavy quarks, despite the fact that the spectrum 
in this case is harder. One last channel could be 
potentially interesting, the direct production of neutrinos,
which is nearly forbidden in the case of neutralinos.
This channel is particularly interesting since in this case 
the spectrum of neutrinos is a line, at energy
equal to the mass of the Kaluza--Klein particle. Rescaling
the fluxes obtained in Ref.~\cite{Bertone:2002ms}
we estimate this flux to be comparable with the
Antares sensitivity to line spectra. A detailed analysis 
of this case will be presented elsewhere. 

\begin{figure}[h]
\includegraphics[width=0.48\textwidth,clip=true]{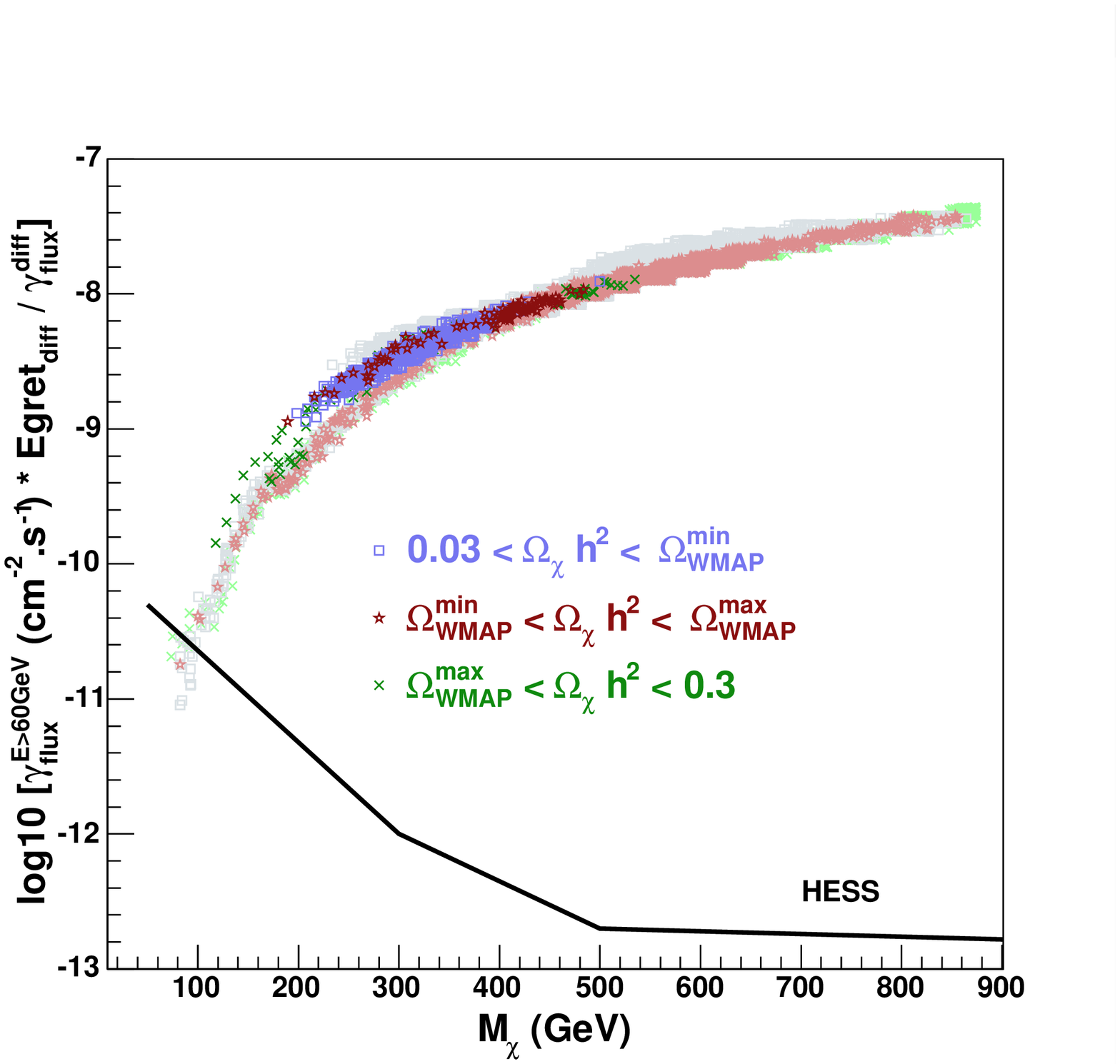}
\caption{Gamma--ray flux (above 60 GeV)  from neutralino annihilation at the 
GC normalized to EGRET. For comparison we show the HESS sensitivity.
Shades paler than in the legend denote a low $\delta_\mu^{\mathrm{susy}}$ value.
}
\label{gam60Egret}
\end{figure}

Finally, to show how our upper bound on the neutrino flux from the Galactic
centre would evolve with new data on gamma ray fluxes, we show in
Fig.~\ref{gam60Egret} the flux above 60~GeV coming from the
same neutralino annihilations in the Galactic centre, applying the same
normalization to EGRET that we used for neutrinos in
Fig.~\ref{NeutrinosEGRETOh2}. As in that figure, the points trace an upper
bound on the gamma flux above 60~GeV, given the EGRET measurement. If Hess
sees a signal (which is not excluded according to Fig.~\ref{gam60Egret}),
{\it e.g.} two orders of magnitudes below this gamma upper bound, the upper
bound on the neutrino flux Fig.~\ref{NeutrinosEGRETOh2} can accordingly be
reduced by two orders of magnitudes.

\section{Conclusions}

The  flux of neutrinos from \DM annihilation at the \GC 
depends on the assumed \DM profile and on the details of 
annihilation of the specific candidate adopted. 
It is nevertheless possible to obtain an upper limit 
for the neutrino flux, by requiring that the associated
gamma-ray emission do not exceed the flux observed by 
EGRET in the direction of the \GC. 

We have estimated such upper limits in the case of neutralinos 
and concluded that any associated neutrino flux lies below 
the experimental sensitivity of
Antares, unless the neutralino mass is above $\sim 700$~GeV . 
In this case, corresponding to models with a low 
$\delta_\mu^{\mathrm{susy}}$ value, and even assuming that 
the gamma-ray emission observed by EGRET is entirely due to 
neutralino annihilation, the upper limit on the neutrino flux
is barely above the minimum signal observable by Antares in 
3 years. 

This means that Antares will not
be able to see neutrinos from neutralino annihilation at the
\GC. Conversely, the positive detection of such a flux would 
either require a different explanation in terms, e.g., of other
astrophysical sources, or the adoption of \DM candidates other 
than neutralinos.

%% \begin{figure}[h]
%% \includegraphics[width=0.48\textwidth,clip=true]{gagaEgret.eps}
%% \caption{Gamma-gamma line  from neutralino annihilation at the 
%% GC normalized to EGRET.}
%% \label{gagalineEgret}
%% \end{figure}

%% \begin{figure}[h]
%% \includegraphics[width=0.48\textwidth,clip=true]{gazEgret.eps}
%% \caption{Gamma-Z line from neutralino annihilation at the 
%% GC normalized to EGRET.}
%% \label{gaZlineEgret}
%% \end{figure}

\section*{Acknowledgements}
It is a pleasure to acknowledge the friendly computer cooperations with Jean-Loïc
Kneur, Genevi\`eve B\'elanger and Yann Mambrini, without whom the code
merging used in this work would not have been possible. GB would like to thank 
G\"unter Sigl for helpful discussions and John Beacom for useful comments.
GB is supported by the DOE and the NASA grant NAG 5-10842 at Fermilab.

\end{document}